\documentclass{article}
\usepackage{amssymb}
\usepackage{amsfonts}
\usepackage{amsmath}
\usepackage{eurosym}

\begin{document}

\title{An Entropic Dynamics Approach to Geometrodynamics\footnote{First presented at the \emph{39th International Workshop on Bayesian Inference and Maximum Entropy Methods in Science and Engineering}.}}
\author{Selman Ipek\footnote{sipek@albany.edu}\hspace{.15 cm} and Ariel Caticha\footnote{acaticha@albany.edu} \\
{\small Physics Department, University at Albany-SUNY, Albany, NY 12222, USA.%
}}
\date{}
\maketitle

\abstract{In the Entropic Dynamics (ED) framework quantum theory is derived as an application of entropic methods of inference. The physics is introduced through appropriate choices of variables and of constraints that codify the relevant physical information.
In previous work, a manifestly covariant ED of quantum scalar fields in a fixed background spacetime was developed. Manifest relativistic covariance was achieved by imposing constraints in the form of Poisson brackets and of intial conditions to be satisfied by a set of local Hamiltonian generators. Our approach succeeded in extending to the quantum domain the classical framework that originated with Dirac and was later developed by Teitelboim and Kuchar. In the present work the ED of quantum fields is extended further by allowing the geometry of spacetime to fully partake in the dynamics. The result is a first-principles ED model that in one limit reproduces quantum mechanics and in another limit reproduces classical general relativity.
Our model shares some formal features with the so-called “semi-classical“ approach to gravity.}









\section{Introduction}
Efforts to develop a theory of quantum gravity (QG) have been dedicated, principally, to two main candidates — string theory (ST) and loop quantum gravity (LQG).  (For a review see e.g., \cite{Review QG}) But despite the general sentiment that these programs are markedly different, they share some fundamental commonalities not usually discussed. Namely, these programs share a rather strict view of QG in which gravitation is itself a quantum force, akin to the electroweak and strong forces of the standard model; such approaches may aptly be referred to as \emph{quantized gravity} theories. But QG, broadly construed, is not simply a program of quantized gravity.  Indeed, although interest in QG is driven by a variety of reasons (see e.g., \cite{Butterfield Isham 2001}), a very basic motivation is that of \emph{consistency}: if matter, which is well-described by quantum theory (QT), is a source of energy, momentum, and so on, and if gravity couples to matter, then there must be some theory, or shared framework that brings the two together. We conjecture that this common thread is, in fact, \emph{entropy}.

We propose here a QG candidate formulated entirely within the Entropic Dynamics (ED) framework. ED is a scheme for generating dynamical theories that are consistent with the entropic and Bayesian rules for processing information.\footnote{For a review of entropic and Bayesian methods, see e.g., \cite{Caticha 2012}\cite{Jaynes Rosenkrantz 1983}.} Among the successes of the ED framework are principled derivations of several aspects of the quantum formalism (For a review of current work, see e.g., \cite{Caticha 2019}) that are also sensitive to, and indeed, help to clarify many conceptual issues that plague QT \cite{Caticha Johnson 2012}. The standout feature of ED that makes this possible is that a clear delineation is maintained between the \emph{ontological}, or physical, aspects of the theory and those that are of \emph{epistemological} significance. The distinction is important: the \emph{ontic}, or physical, variables are the subject of our inferences, and it is their values we wish to predict. The ability to make such predictions, however, depends on the available information, which in ED is codified in the constraints. In short, ED is a dynamics driven by constraints.

In previous work \cite{Ipek et al 2017}\cite{Ipek et al 2018}, ED was utilized to derive the standard quantum field theory on curved space-time (QFTCS) by making a particular choice of constraints, appropriate for a scalar field $\chi_{x}$ propagating on a \emph{fixed} background space-time. One constraint involved the introduction of a drift potential $\phi$ that guides the flow of the probability $\rho$, while another was the employment of a canonical formalism, i.e. \emph{Hamiltonians}, for driving their joint dynamics. But in the context of a manifestly relativistic theory, further constraints are needed. In \cite{Ipek et al 2017}\cite{Ipek et al 2018} these were supplied by the adoption of  the covariant canonical methods of Dirac, Hojman, Kucha\v{r}, and Teitelboim (DHKT). We briefly review their work. Within the DHKT scheme, time evolution unfolds as an accumulation of local deformations in three-dimensional space, constrained by the requirement that the evolution of three-dimensional space be such that it sweeps a four-dimensional space-time. The criterion for accomplishing this, called \emph{path independence} by Kucha\v{r} and Teitelboim, results in an ``algebra" to be satisfied by the generators of deformations.\footnote{The quotes in ``algebra" are meant as a reminder that local deformations do not form a true group; while two deformations can be composed to form another, the composition depends on the original surface.} An interesting aspect of the DHKT approach is that whether one deals with an externally prescribed space-time, or whether the background geometry is itself truly dynamical, the ``algebra" to be satisfied remains the \emph{same} \cite{Teitelboim 1972} (see also \cite{Kuchar 1973}). The true distinction manifests in the choice of variables to describe the evolving geometry.

Our goal here is to pursue an ED in which the background is itself a full partner in the dynamics. But we also wish to proceed conservatively; our aim is not,after all, to simply discard information which has already been proven valuable in ED. Thus we proceed in a minimalist fashion by taking the constraints of \cite{Ipek et al 2017}\cite{Ipek et al 2018} and altering them to account for the additional information; which amounts here to a different choice of \emph{variables} to describe the geometry. Following the seminal work of Hojman, Kucha\v{r}, and Teitelboim in \cite{Hojman Kuchar Teitelboim 1976}, we make a choice in which the metric of three-dimensional space is itself a canonical variable. The ED that emerges from this choice bears great formal resemblance to the so-called ``semi-classical" Einstein equations, but the conceptual differences are enormous: (a) in direct contradiction to the standard (Copenhagen) interpretation of QT, the scalar matter field is ontic and thus has definite values at all times. (b) There are no quantum probabilities. The ED approach is in strict adherence to the Bayesian and entropic principles of inference. (c) Indeed, because of the abidance to the Bayesian view of probability, there are none of the paradoxes associated to the quantum measurement problem --- which is the main source of objection to semi-classical gravity (see \cite{Pro Quantize} and references therein). (d) Our geometrodynamics is not a forced marriage between classical gravity and quantum field theory. It is a first principles framework that not only derives the correct coupling of matter to gravity, but also reconstructs the theory of quantum mechanics itself.

The paper is outlined as follows. In section \ref{ED_Steps}, we give a quick review of the ED of infinitesimal steps, while we review some basic space-time kinematics in section \ref{ST_kinematics}. In section \ref{ED_time} we introduce the notion of a relativistic notion of entropic time. Many of the new results are in sections \ref{ED_GMD} and \ref{ED_QT}, where we construct a geometrodynamics driven by entropic matter. We discuss our results in section \ref{conclusion}.

\section{Reviewing the Entropic Dynamics of infinitesimal steps}\label{ED_Steps}
We adopt the notations and conventions of \cite{Ipek et al 2017}\cite{Ipek et al 2018} throughout. We study a single scalar field $\chi \left( x\right)\equiv \chi_{x} $ whose values are posited to be definite, but unknown. An entire field configuration, denoted $\chi $, lives on a $3$-dimensional curved space $\sigma $, the points of which are labeled by coordinates $x^{i}$ ($i=1,2,3$). The space $\sigma$ is a three-dimensional curved space equipped with a metric $g_{ij}$ that is currently fixed, but that will later become dynamical. A single field configuration $\chi$ is a point in an $\infty $-dimensional configuration space $\mathcal{C}$. Our uncertainty in the values of the field is then given by a probability distribution $\rho \lbrack \chi ]$ over $\mathcal{C}$, so that the proper probability that the field attains a value $\hat{\chi}$ in an infinitesimal region of $\mathcal{C}$ is $\text{Prob}[\chi < \hat{\chi} < \chi+\delta\chi]=\rho[\chi]\, D\chi$, where $D\chi$ is an integration measure over $\mathcal{C}$.
\vskip .25 cm
\noindent \textbf{Maximum Entropy---} The goal is to predict the evolution of the scalar field $\chi $. To this end we make one major assumption: in ED, the fields follow continuous trajectories such that finite changes can be analyzed as an accumulation of many infinitesimally small ones. Thus we are interested in obtaining the probability $P\left[ \chi ^{\prime }|\chi \right] $ of a transition from an initial configuration $\chi $ to a neighboring $\chi ^{\prime }=\chi +\Delta \chi $. This is accomplished via the Maximum Entropy (ME) method by maximizing the entropy functional, 
\begin{eqnarray}
S\left[ P,Q\right] =-\int D\chi ^{\prime }P\left[ \chi ^{\prime }|\chi %
\right] \log \frac{P\left[ \chi ^{\prime }|\chi \right] }{Q\left[ \chi
^{\prime }|\chi \right] },  \label{entropy a}
\end{eqnarray}
relative to a prior $Q\left[ \chi ^{\prime }|\chi \right] $ and subject to appropriate constraints.

The prior $Q\left[ \chi ^{\prime }|\chi \right] $ that incorporates
the information that the fields change by infinitesimally small amounts, but which is otherwise maximally uninformative is a product of Gaussians,\footnote{This prior can itself be derived from the ME method, and in that case, the $\alpha_{x}$ appear as Lagrange multipliers.} 
\begin{eqnarray}
Q\left[ \chi ^{\prime }|\chi \right] \propto \,\exp -\frac{1}{2}\int
dx\,g_{x}^{1/2}\alpha _{x}\left( \Delta \chi _{x}\right) ^{2}~  \label{prior}
\end{eqnarray}%
where $g_{x}^{1/2} = \det|g_{ij}|^{1/2}$ is a scalar density of weight one. (For notational simplicity we write $dx^{\prime }$ instead of $d^{3}x^{\prime }$.) Continuity is enforced in the limit that the scalar-valued parameters $\alpha_{x}\to\infty$. As argued in  \cite{Ipek et al 2017}\cite{Ipek et al 2018}, a single additional constraint is required to develop a richer quantum dynamics,\footnote{Note that since $\chi _{x}$ and $\Delta \chi _{x}$ are scalars, in order that (\ref{Constraint 2}) be invariant under coordinate transformations of the surface the derivative $\delta /\delta \chi _{x}$ must transform as a scalar density.}
\begin{eqnarray}
\langle \Delta \phi \rangle = \int_{%
\mathcal{C}}D\chi ^{\prime }\,P\left[ \chi ^{\prime }|\chi \right] \int
dx\,\,\Delta \chi _{x}\frac{\delta \phi \left[ \chi \right] }{\delta \chi
_{x}}=\kappa ^{\prime },  \label{Constraint 2}
\end{eqnarray}%
which involves the introduction of a
\textquotedblleft drift\textquotedblright\ potential $\phi \lbrack \chi ]$ whose complete justification is still a subject of future investigation.\footnote{There is strong reason to believe more compelling answers will come from the ED of Fermions.} Maximizing (\ref{entropy a}) subject to (\ref{Constraint 2}) and normalization, we obtain a Gaussian transition probability distribution, 
\begin{eqnarray}
P\left[ \chi ^{\prime }|\chi \right] =\frac{1}{Z\left[ \alpha _{x},g_{x}%
\right] }\,\exp -\frac{1}{2}\int dx\,g_{x}^{1/2}\alpha _{x}\left( \Delta
\chi _{x}-\frac{1}{g_{x}^{1/2}\alpha _{x}}\frac{\delta \phi \left[ \chi %
\right] }{\delta \chi _{x}}\right) ^{2},  \label{Trans Prob}
\end{eqnarray}%
where $Z\left[ \alpha _{x},g_{x}\right] $ is the normalization constant. The Gaussian form of (\ref{Trans Prob}) allows us to present a generic change, 
\begin{eqnarray}
\Delta \chi _{x}=\left\langle \Delta \chi _{x}\right\rangle +\Delta w_{x}~,
\end{eqnarray}
as resulting from an expected drift $\left\langle \Delta \chi
_{x}\right\rangle $ plus fluctuations $\Delta w_{x}$. While $\left \langle \Delta w_{x} \right \rangle = 0$, because the distribution is Gaussian, the square fluctuations and expected short steps do not. That is,
\begin{eqnarray}
\left\langle \Delta w_{x}\Delta w_{x^{\prime }}\right\rangle =\frac{1}{%
g_{x}^{1/2}\alpha _{x}}\delta _{xx^{\prime }} \quad \text{and} \quad \left\langle \Delta \chi _{x}\right\rangle =\frac{1}{g_{x}^{1/2}\,\alpha _{x}%
}\frac{\delta \phi \left[ \chi \right] }{\delta \chi _{x}}\equiv \Delta \bar{%
\chi}_{x}~.   \label{Exp Step 1}
\end{eqnarray}%
Therefore, as in \cite{Ipek et al 2017}\cite{Ipek et al 2018}, the fluctuations dominate the trajectory leading to a Brownian motion.

\section{Some space-time kinematics}\label{ST_kinematics}
In geometrodynamics, the primary object of interest is the three-dimensional metric of space $g_{ij}$, whose time evolution is required to sweep a four-dimensional space-time. Thus coordinates $X^{\mu}$ ($\mu ,\nu ,...$ $=0,1,2,3$) can still be assigned to space-time, even if its geometry \emph{a priori} unknown. By construction, space-time can be foliated by a sequence of space-like surfaces $\left\{ \sigma \right\} $; that is, its topology is globally hyperbolic. The embedding of a surface in space-time is defined by four embedding functions $X^{\mu }=X^{\mu }\left(x^{i}\right) $. 

An infinitesimal deformation of the surface $\sigma $ to a neighboring surface $\sigma^{\prime }$ is determined by the deformation vector 
\begin{eqnarray}
\delta \xi ^{\mu }=\delta \xi ^{\bot }n^{\mu }+\delta \xi ^{i}X_{i}^{\mu }~,
\label{deformation vector}
\end{eqnarray}%
where $n^{\mu }$ is the unit normal to the surface ($n_{\mu }n^{\mu }=-1$, $n_{\mu }X_{i}^{\mu }=0$) and where $X_{i}^{\mu}(x) =\partial_{ix}X^{\mu}(x) $ are the space-time components of vectors tangent to the surface. The normal and tangential components of $\delta \xi ^{\mu }$, also known as the lapse and shift, are collectively denoted ${}(\delta \xi
^{\bot },\delta \xi ^{i})=\delta \xi ^{A}$ and are given by%
\begin{eqnarray}
\delta \xi _{x}^{\bot }=-n_{\mu x}\delta \xi _{x}^{\mu }\quad \text{and}%
\quad \delta \xi _{x}^{i}=X_{\mu x}^{i}\delta \xi _{x}^{\mu }~,
\end{eqnarray}%
where $X_{\mu x}^{i}=g^{ij}g_{\mu \nu }X_{jx}^{\nu }$.
\section{Entropic time}\label{ED_time}
In ED, entropic time is introduced as a tool for keeping track of the accumulation of many short steps. (For additional details on entropic time, see e.g., \cite{Caticha 2012}.) Here we introduce a manifestly covariant notion of entropic time, along the lines of that in \cite{Ipek et al 2017}\cite{Ipek et al 2018}.
\vskip .25 cm
\noindent \textbf{Ordered instants---}
Central to our formulation of entropic time is the notion of an instant of time. In a properly relativistic theory in curved space-time, such a notion is provided by an \emph{arbitrary} space-like surface, denoted $\sigma$ (see e.g., \cite{Kuchar 1973}). This allows us to define the epistemic state at the instant $\sigma$, characterized by the probability $\rho_{\sigma}[\chi]$, the drift potential $\phi_{\sigma}[\chi]$, etc.

Having established the notion of an instant, including a assignment of the instantaneous probability $\rho_{\sigma}[\chi]$, our task turns to updating from one instant to the next. Such dynamical information is encoded in the short-step transition probability from eq.(\ref{Trans Prob}), or better yet, the joint probability $ P\left[ \chi ^{\prime },\chi \right]= P\left[ \chi ^{\prime }|\chi \right]\rho_{\sigma}[\chi] $. A straightforward applications of the ``sum rule" of probability theory suggests that the probability at the next instant is given by
\begin{eqnarray}
\rho_{\sigma^{\prime}}[\chi^{\prime}] = \int D\chi \, P\left[ \chi ^{\prime }|\chi \right]\rho_{\sigma}[\chi]~.\label{CK equation}
\end{eqnarray}
This is the basic dynamical equation for the evolution of probability.
\vskip .25 cm
\noindent \textbf{Duration---}
To complete our construction of time we must specify the duration between instants. In ED time is defined so that motion looks simple. Since for short steps the dynamics is dominated by fluctuations, eq.(\ref{Exp Step 1}), the specification of the time interval is achieved through an appropriate choice of the multipliers $\alpha _{x}$. Moreover, following \cite{Ipek et al 2017}\cite{Ipek et al 2018}, since we deal here with the duration between curved spaces, this notion of separation should be local, and it is natural to define duration in terms of the local proper time $\delta_{x}^{\perp}$. More specifically, let 
\begin{eqnarray}
\alpha _{x}=\frac{1}{ \delta \xi _{x}^{\bot }}\quad \text{so that}\quad
\left\langle \Delta w_{x}\Delta w_{x^{\prime }}\right\rangle =\frac{
\,\delta \xi _{x}^{\bot }}{g_{x}^{1/2}}\delta _{xx^{\prime }}~.
\label{Duration}
\end{eqnarray}%

\vskip .25 cm
\noindent \textbf{The local-time diffusion equations---}
The dynamics expressed in integral form by (\ref{CK equation}) and (%
\ref{Duration}) can be rewritten in differential form as an infinite set of local equations, one for each spatial point,
\begin{eqnarray}
\frac{\delta \rho _{\sigma }}{\delta \xi _{x}^{\bot }}=-\,g_{x}^{-1/2}\frac{%
\delta }{\delta \chi _{x}}\left( \rho _{\sigma }\,\frac{\delta \Phi _{\sigma
}}{\delta \chi _{x}}\right)\quad\text{with}\quad \Phi
_{\sigma }\left[ \chi \right] = \,\phi _{\sigma }\left[ \chi \right]
- \log \rho _{\sigma }^{1/2}\left[ \chi \right] ~. \label{FP equation}
\end{eqnarray}%
As shown in \cite{Ipek et al 2017}\cite{Ipek et al 2018}, this set of equations describes the flow of the probability $\rho _{\sigma }\left[ \chi \right] $ in the configuration space $\mathcal{C}$ in response to the geometry and the functional $\Phi_{\sigma}[\chi]$, which will eventually be identified as the Hamilton-Jacobi functional, or the phase of the wave functional in the quantum theory. Moreover, when a particular foliation is chosen, these equations collectively form a Fokker-Planck equation, thus justifying the name, the Local-Time Fokker-Planck (LTFP) equations for eqns.(\ref{FP equation}).
\section{Geometrodynamics driven by entropic matter}\label{ED_GMD}
In an \textit{entropic} dynamics, evolution is driven by information codified into constraints. An entropic geometrodynamics, it follows, consists of dynamics driven by a specific choice of constraints, which we discuss here. In \cite{Ipek et al 2017}\cite{Ipek et al 2018}, QFTCS was derived under the assumption that the background remains fixed. But such assumptions, we know, should break down when one considers states describing a non-negligible concentration of energy and momentum. Thus we must revise our constraints appropriately. A natural way to proceed is thus to allow the geometry itself to take part in the dynamical process: the geometry affects $\rho_{\sigma}[\chi]$ and $\phi_{\sigma}[\chi]$, they then act back on the geometry, and so forth. Our goal here is to make this interplay concrete.
\vskip .25 cm
\noindent \textbf{Path independence---} In a relativistic theory there are many ways to evolve from an initial instant to a final one, and each way must agree. This is the basic insight by DHKT in their development of manifestly covariant dynamical theories. The implementation of this idea, through the principle of \emph{path independence}, leads to a set of Poisson brackets (see e.g., \cite{Teitelboim 1972})
\begin{eqnarray}
\left\{ H_{\bot x},H_{\bot x^{\prime }}\right\} &=&(g_{x}^{ij}H_{jx}+g_{x^{\prime
}}^{ij}H_{jx^{\prime }})\partial _{ix}\delta (x,x^{\prime })~,  \label{PB 1}
\\
\left\{ H_{ix},H_{\bot x^{\prime }}\right \} &=&H_{\bot x}\partial _{ix}\delta
(x,x^{\prime })~,  \label{PB 2} \\
\left\{ H_{ix},H_{jx^{\prime }}\right \} &=&H_{ix^{\prime }}\,\partial _{jx}\delta
(x,x^{\prime })+H_{jx}\,\partial _{ix}\delta (x,x^{\prime })~,  \label{PB 3}
\end{eqnarray}%
supplemented by the constraints
\begin{eqnarray}
H_{\perp x} \approx 0\quad\text{and}\quad H_{ix} \approx 0 ~,\label{Hamiltonian constraints}
\end{eqnarray}
where ``$\approx$" is understood as a \textit{weak} equality \cite{Bergmann 1949}. (From a practical viewpoint, the Poisson bracket relations are essentially constraints on the allowed functional form of the generators $H_{Ax}$ for arbitrary choices of the dynamical variables. On the other hand, the weak constraints $H_{Ax}\approx 0$ are meant to restrict the allowed choices of initial conditions for a given form of the generators $H_{Ax}$.)
\vskip .25 cm
\noindent \textbf{The phase space---} The equations (\ref{PB 1})-(\ref{Hamiltonian constraints}) of path independence are \emph{universal}. That is, if the dynamics is to be relativistic, these equations must hold. Whatever the choice of canonical variables, or whether the background is fixed or dynamical, the same ``algebra" must hold.

As one might expect, the ED formulated here with a dynamical background shares some similarities with the theory developed in \cite{Ipek et al 2017}\cite{Ipek et al 2018}, in the context of a fixed background. Most obvious is that the variables $\rho$ and $\phi$, or equivalently, $\rho$ and $\Phi$ remain canonically conjugate, forming the so-called \emph{ensemble phase space}, or e-phase space for short. However a crucial difference emerges with respect to the treatment of the geometry. Here, deviating from \cite{Ipek et al 2017}\cite{Ipek et al 2018}, we instead follow Hojman, Kucha\v{r}, and Teitelboim (HKT) by describing the dynamics of the geometry by taking the metric $g_{ij}(x)$ as a central dynamical variable. Of course, for the scheme to be canonical, we must also introduce a set of auxiliary variables $\pi^{ij}$, which must have the character of tensor densities. At this juncture the sole interpretation of the $\pi^{ij}$ are as the momenta conjugate to $g_{ij}$, defined by the canonical Poisson bracket relations
\begin{eqnarray}
\left\{g_{ij}(x),g_{kl}(x^{\prime})\right \} &=& \left\{\pi^{ij}(x),\pi^{kl}(x^{\prime})\right \} = 0~,\notag\\
\left\{g_{ij}(x),\pi^{kl}(x^{\prime})\right \}  &=& \frac{1}{2}\left(\delta^{k}_{i}\delta^{l}_{j}+\delta^{k}_{j}\delta^{l}_{i}\right)\delta(x,x^{\prime})~.\label{canonical relations Grav}
\end{eqnarray}
Here we have introduced the notion of a Poisson bracket, which is an anti-symmetric bi-linear product that allows for the notion of an algebra. Written in local coordinates, the Poisson brackets take the form
\begin{eqnarray}
\left\{F,G\right\} &=& \int dx\left(\frac{\delta F}{\delta g_{ij}(x)}\frac{\delta G}{\delta \pi^{ij}(x)}-\frac{\delta G}{\delta g_{ij}(x)}\frac{\delta F}{\delta \pi^{ij}(x)}\right)\notag\\
&+&\int D\chi\left(\frac{\tilde{\delta} F}{\tilde{\delta}\rho[\chi] }\frac{\tilde{\delta} G}{\tilde{\delta} \Phi[\chi]}-\frac{\tilde{\delta} G}{\tilde{\delta}\rho[\chi] }\frac{\tilde{\delta} F}{\tilde{\delta} \Phi[\chi]}\right)~,
\end{eqnarray}
for arbitrary functionals $F$ and $G$ of the phase space variables $(\rho,\Phi;g_{ij},\pi^{ij})$.



\vskip .25 cm
\noindent \textbf{The super-momentum---} We now turn our attention to the local Hamiltonian generators $H_{Ax}$, and more specifically, we look to provide explicit expressions for these generators in terms of the canonical variables by solving the Poisson brackets eqns.(\ref{PB 1})-(\ref{PB 3}). We begin with the tangential generator $H_{ix}$, which generates changes in the canonical variables by dragging them parallel to the space $\sigma$. As shown in \cite{Teitelboim 1972}\cite{Hojman Kuchar Teitelboim 1976}, the tangential generator can be shown to split
\begin{eqnarray}
H_{Ax}[\rho,\Phi;g_{ij},\pi^{ij}] = H^{G}_{Ax}[g_{ij},\pi^{ij}]+\tilde{H}_{Ax}[\rho,\Phi],\label{Super-Momentum split}
\end{eqnarray}
into components we identify as an ensemble super-momentum $\tilde{H}_{ix}$ and a gravitational super-momentum $H^{G}_{ix}$. This then leads to eq.(\ref{PB 3}) similarly decomposing into ensemble and gravitational pieces so that each can be solved independently of the other. The appropriate super-momentum for the ensemble sector was given in \cite{Ipek et al 2017}\cite{Ipek et al 2018}, with the result that
\begin{eqnarray}
\tilde{H}_{ix} = - \int D\chi \rho[\chi]\frac{\delta\Phi[\chi]}{\delta\chi_{x}}\partial_{ix}\chi_{x}~, \label{Super momentum}
\end{eqnarray}
while
\begin{eqnarray}
\quad H_{ix}^{G} = -2 \partial_{j}\left(\pi^{jk}\, g_{ik}\right) + \pi^{jk}\partial_{i}g_{jk} 
\end{eqnarray}
is the gravitational contribution, determined by Hojman et al in \cite{Hojman Kuchar Teitelboim 1976}. The so-called super-momentum constraint from eq.(\ref{Hamiltonian constraints}) is then just
\begin{eqnarray}
H_{ix} =  -2 \partial_{j}\left(\pi^{jk}\, g_{ik}\right) + \pi^{jk}\partial_{i}g_{jk} - \int D\chi \rho[\chi]\frac{\delta\Phi[\chi]}{\delta\chi_{x}}\partial_{ix}\chi_{x}\approx 0~.\label{Super-momentum constraint explicit}
\end{eqnarray}
\vskip .25 cm
\noindent \textbf{The super-Hamiltonian---} As pertains to the super-Hamiltonian, a similar decomposition does not occur. But following Teitelboim \cite{Teitelboim 1972} let us suggestively rewrite $H_{\perp x}$ as
\begin{eqnarray}
H_{\perp x} = H_{\perp x}^{G}[g_{ij},\pi^{ij}]+\tilde{H}_{\perp x}[\rho,\Phi;g_{ij}, \pi^{ij}]~.\notag
\end{eqnarray}
Note we make no assumptions in writing $H_{\perp x}$ in this way, as this simply \emph{defines} the contribution of ``matter." The assumption comes, instead, the requirement that the ensemble super-Hamiltonian $\tilde{H}_{\perp x}$ to be \emph{independent} of the momentum variable $\pi^{ij}$.\footnote{Although many interesting models remain after this assumption, some models of physical interest are, indeed, excluded by this simplification. We leave it to future work to relax this requirement.} (This simplification is called the assumption of ``non-derivative" coupling, since it can be \emph{proven} \cite{Teitelboim 1972} that this implies $\tilde{H}_{\perp x}$ is just a local function (not functional) of $g_{ij}$, not its derivatives.) Another consequence of this assumption is that eq.(\ref{PB 1}) --- the most difficult Poisson bracket --- decomposes completely into gravitational and matter sectors. Thus each can be approached independently. That is, the gravitational side can proceed \emph{as if} there were no sources, while the matter side can proceed along lines similar to \cite{Ipek et al 2017}\cite{Ipek et al 2018}.

The solution to the gravitational piece, which is quite involved, was first given by HKT \cite{Hojman Kuchar Teitelboim 1976} and takes the form\footnote{Note that the solution given in \cite{Hojman Kuchar Teitelboim 1976} relies on the assumption that geometrodynamics is time-reversible. An alternative derivation in \cite{Kuchar 1973} obtains the same result without this assumption, but uses a Lagrangian instead.}
\begin{eqnarray}
H_{\perp x}^{G} = \kappa \, G_{ijkl}\pi^{ij}\pi^{kl} - \frac{g^{1/2}}{2\kappa}R~,
 \label{Super hamiltonian grav}
\end{eqnarray}
where
\begin{eqnarray}
G_{ijkl} =  g^{-1/2}\left(g_{ik}g_{jl}+g_{il}g_{jk}-g_{ij}g_{kl}\right)\label{DeWitt super metric}
\end{eqnarray}
is the DeWitt super metric, $R$ is the Ricci scalar for three-dimensional space, and $\kappa$ is a constant coefficient. (We have set the cosmological constant $\lambda = 0$.) The determination of the ensemble super-Hamiltonian is subject not only to satisfying the Poisson bracket eq.(\ref{PB 1}), but also to the requirement that the evolution generated by $\tilde{H}_{\perp x}$ reproduces the LTFP eqns.(\ref{FP equation}). In \cite{Ipek et al 2017}\cite{Ipek et al 2018} it was shown that an acceptable (but not exhaustive) family of ensemble super-Hamiltonians is given by
\begin{eqnarray}
\tilde{H}_{\perp x}[\rho,\Phi] = \tilde{H}_{\perp x}^{0}[\rho,\Phi;g_{ij}]+F_{x}[\rho; g_{ij}]\label{Super Hamiltonian F}
\end{eqnarray}
with
\begin{eqnarray}
\tilde{H}_{\perp x}^{0}[\rho,\Phi] = \frac{1}{2}\int D\chi \rho\left(\frac{1}{g_{x}^{1/2}}\left(\frac{\delta\Phi}{\delta\chi_{x}}\right)^{2}+g_{x}^{1/2}g^{ij}(x)\partial_{ix}\chi_{x}\partial_{jx}\chi_{x}\right)~,\label{Super hamiltonian free}
\end{eqnarray}
where the functional $F_{x}[\rho;g_{ij}]$ is restricted to the simple form
\begin{eqnarray}
F_{x}[\rho;g_{ij}] = \int D\chi \, \rho \left (g^{1/2}_{x}V_{x}(\chi_{x}) + \frac{\beta}{g^{1/2}_{x}} \left ( \frac{\delta\log\rho}{\delta\chi_{x}}\right )^{2}\right )~,
\end{eqnarray}
where the potential $V_{x}(\chi)$ is a function only of the field and $\beta$ is a coupling constant. For future convenience we set to $\beta = 1/8$.

From eqns.(\ref{Super hamiltonian grav}) and (\ref{Super Hamiltonian F}) the super-Hamiltonian constraint is then just
\begin{eqnarray}
H_{\perp x} = H^{G}_{\perp x}+\tilde{H}_{\perp x} \approx 0~,\label{Super-Hamiltonian constraint split}
\end{eqnarray}
where the gravitational and ensemble pieces are those given in eqns.(\ref{Super hamiltonian grav}) and (\ref{Super Hamiltonian F}). Note that a solution of this constraint requires fixing a set of variables in terms of another set --- i.e. the gravitational variables are not necessarily independent of the probability $\rho$ and phase $\Phi$! 
\section{The dynamical equations}\label{ED_QT}
In the previous section we have identified a representation of the relations eqns.(\ref{PB 1})-(\ref{Hamiltonian constraints}) in terms of the canonical variables $(\rho,\Phi;g_{ij},\pi^{ij})$. The resulting evolution, generated by these $H_{Ax}$, leads to a fully covariant geometrodynamics driven by entropic matter. But to do this, we first pick a foliation of space-time with parameter $t$, specified by a particular choice of \emph{lapse} and \emph{shift} functions, which are, respectively, given by
\begin{eqnarray}
N(x,t) = \frac{\delta\xi^{\perp}_{x}}{dt}\quad\text{and}\quad N^{i}(x,t) = \frac{\delta\xi^{i}_{x}}{dt}~.
\end{eqnarray}
\vskip .25 cm
\noindent \textbf{The Schr\"{o}dinger equation---} We are interested in the dynamical evolution of the ensemble variables $\rho$ and $\Phi$, however, this very same dynamics can be expressed equivalently by the introduction of complex variables $\Psi_{t} = \rho^{1/2}_{t}e^{i\Phi_{t}}$ and $\Psi_{t}^{*} = \rho^{1/2}_{t}e^{-i\Phi_{t}}$ (we use units in which $\hbar = 1$). The reason these variable turn out to be useful, is that the dynamical equations turn out to take a familiar form. In particular, we have
\begin{eqnarray}
i\partial_{t}\Psi_{t}[\chi] &=& i\int dx\left[ \left\{\Psi_{t}[\chi],H_{\perp x}\right \}N(x,t)+\left\{\Psi_{t}[\chi],H_{i x}\right \}N^{i}(x,t)\right]\notag\\
&=& \int dx\left [N(x,t)\hat{H}_{\perp x}+N^{i}(x,t)\hat{H}_{ix}\right ]\Psi_{t}[\chi]\label{Schrodinger equation}~,
\end{eqnarray}
where $\hat{H}_{ix}$ and $\hat{H}_{\perp x}$ are given by
\begin{eqnarray}
\hat{H}_{ix} = i\,\partial_{i}\chi_{x}\frac{\delta}{\delta\chi_{x}}
\label{Super momentum operator}
\end{eqnarray}
and
\begin{eqnarray}
\hat{H}_{\perp x} =- \frac{1}{2g^{1/2}}\frac{\delta^{2}}{\delta\chi^{2}_{x}}+\frac{g^{1/2}}{2}g^{ij}\partial_{i}\chi_{x}\partial_{j}\chi_{x}+g^{1/2}V_{x}(\chi_{x};g_{ij})~,\label{Super Hamiltonian operator}
\end{eqnarray}
respectively. Notice that although eq.(\ref{Schrodinger equation}) is ostensibly a linear equation for the functional $\Psi_{t}$, which may suggest calling this a ``Schr\"{o}dinger equation," closer inspection reveals this to be misleading. Indeed, owing to the constraint eqns.(\ref{Super-momentum constraint explicit}) and (\ref{Super-Hamiltonian constraint split}), the metric $g_{ij}$ that appears in this equation itself depends on the variables $\Psi$ and $\Psi^{*}$, leading instead to a \emph{non-linear} equation!
\vskip .25 cm
\noindent \textbf{Geometrodynamics---} To complete the description of the dynamics we will determine the evolution of the geometrical variables $(g_{ij},\pi^{ij})$. Beginning with the metric, its time evolution, generated by the super-Hamiltonians $H_{Ax}$ given above, after a straightforward computation yields 
\begin{eqnarray}
\partial_{t}g_{ij} =  \frac{2\kappa N(x,t)}{g_{x}^{1/2}}\left(2\pi_{ij}(x) - \pi(x) g_{ij}(x)\right)+\nabla_{i}N_{j}(x,t)+\nabla_{j}N_{i}(x,t),\label{Evolution equation metric}
\end{eqnarray}
where $\nabla_{i}$ is the metric compatible covariant derivative.\footnote{Note that eq.(\ref{Evolution equation metric}) relates the conjugate momentum $\pi^{ij}$ to the extrinsic curvature of the surface $K_{ij}$.} The equation for the conjugate momentum $\pi^{ij}$ is more interesting
\begin{eqnarray}
\partial_{t}\pi^{ij}  &=& -\frac{Ng^{1/2}}{2\kappa}\left(R^{ij}-\frac{1}{2}g^{ij}R\right )+ \frac{N\kappa}{g^{1/2}}g^{ij}\left(\pi^{kl}\pi_{kl}-\frac{1}{2}\pi^{2}\right)\notag\\
&-&4\frac{N\kappa}{g^{1/2}}\left(\pi^{ik}\pi^{j}_{k}-\frac{1}{2}\pi\pi^{ij}\right)+\frac{g^{1/2}}{2\kappa}\left(\nabla^{i}\nabla^{j}N - g^{ij}\nabla^{k}\nabla_{k}N\right)+\nabla_{k}\left(\pi^{ij}N^{k}\right)\notag\\
&-& \pi^{ik}\nabla_{k}N^{j} - \pi^{kj}\nabla_{k}N^{i}-N\frac{\partial\tilde{H}_{\perp x}}{\partial g_{ij}}~,\label{Evolution equation pi}
\end{eqnarray}
where we have introduced $R^{ij}$, the Ricci tensor.\footnote{We have used the fact that the non-derivative coupling assumption implies that $\tilde{H}_{\perp x}$ is local in $g_{ij}$.} Note that the evolution of $\pi^{ij}$ is driven by $\partial \tilde{H}_{\perp x}/\partial g_{ij}$, which is a functional of $\Psi$. Thus the evolution eqns.(\ref{Evolution equation metric}) and (\ref{Evolution equation pi}) describe a queer dynamics in which the geometry evolves in response to the informational state, codified by the functional $\Psi$. This will be further explored below.

\section{Conclusions and discussion}\label{conclusion}
The ED developed here has several interesting features. Although written in the relatively less common language of geometrodynamics, the evolution eqns.(\ref{Evolution equation metric}) and (\ref{Evolution equation pi}), together with the constraints (\ref{Super-momentum constraint explicit}) and (\ref{Super-Hamiltonian constraint split}) are mathematically equivalent to the so-called ``semi-classical" Einstein equations (SCEE), which are typically written as \cite{Butterfield Isham 2001}
\begin{eqnarray}
G_{\mu\nu} =8\pi \kappa \left \langle \hat{T}_{\mu\nu} \right \rangle~,\label{Semi classical equations}
\end{eqnarray}
where $G_{\mu\nu}$ is the Einstein tensor and $\left\langle \hat{T_{\mu\nu}}\right \rangle$ is the expected value of energy-momentum operator of a quantum scalar field.

Such a theory of gravity has long been seen as a desirable step intermediate to a full theory of QG, in part because it contains well-established physics --- QFTCS and classical general relativity --- in the limiting cases where they are valid. But there has been much debate (see e.g., \cite{Pro Quantize}), on the other hand, as to the status of semi-classical theories as true QG candidate; with many harboring a negative view.

Here we do not propose a definitive rebuke of all these critics, but note that the ED formulation of SCEE has certain features that allow it to evade the most cogent criticisms. For one, a problem that is often raised against the SCEE is that it is proposed in a rather \emph{ad hoc} manner, based on heuristic arguments. But in ED these equations are \emph{derived} on the basis of well-defined assumptions and constraints. Period.

Another argument commonly raised against the SCEE (see e.g., the paper by Unruh in \cite{Pro Quantize}) is that the left hand side of eq.(\ref{Semi classical equations}) features the gravitational field, which is commonly viewed as a real ``physical" field, while the right hand side contains the quantum state $\Psi$, which is an epistemic variable that describes the known information. Taken at face value, this leads to serious issues during the process of measurement and subsequent wave function ``collapse." As mentioned by Kibble \cite{Pro Quantize}, however, the standard theory of measurement in QT is itself fraught with issues and thus serves as a weak basis on which to judge the merits of the SCEE. The ED approach, on the other hand, allows for a more cogent resolution of the so-called ``measurement problem" of QT \cite{Caticha Johnson 2012} which helps to sidestep many of the standard problems. Moreover, the very premise that the gravitational field be treated as a physical entity is itself dubious. Where, after all, do metrics come from? Who ordered that? One possible answer is offered in \cite{Caticha 2016}, where it is argued that the geometry of space is itself an \emph{information geometry} --- i.e. that spatial distances measure the distinguishability between probability distributions. Certainly such a model places the gravitational field in a far different light than the classical picture typically used to interpret the SCEE.

Finally, the Schr\"{o}dinger equation we obtain here is quite unorthodox. As mentioned above, the dynamics of $\Psi$ follows a \emph{non-linear} equation. This non-linearity is not, however, an artifact of a bad approximation, but the prediction of a theory derived from first principles. This begs the question, is linearity just a misguided prejudice? Or, will the superposition principle become the first casualty of quantum gravity?

\end{document}